\documentclass[aps,pra,reprint,floatfix,showpacs,twocolumn]{revtex4-2}
\usepackage{amsmath}
\usepackage[colorlinks=true,citecolor=blue,linkcolor=blue,urlcolor=blue]{hyperref}
\usepackage[usenames,dvipsnames]{xcolor}
\usepackage{longtable}
\usepackage{graphicx}
\usepackage{color}
\usepackage{float}
\usepackage{placeins}
\usepackage{rotating}
\usepackage{epsfig,rotate}
\usepackage{soul}
\usepackage{multirow}
\usepackage{booktabs}
\usepackage{adjustbox}

\begin{document}

	\title{Dielectronic recombination studies of ions relevant to kilonovae and non-LTE plasma}

	\author{S.~Singh}
	\email{suvam.singh@mpi-hd.mpg.de}
	\author{Z.~Harman}
	\author{C.~H.~Keitel}

	\affiliation{Max--Planck--Institut f\"{u}r Kernphysik, Saupfercheckweg 1, 69117 Heidelberg, Germany}

	\begin{abstract}	
		
		This study presents calculations of rate coefficients, resonance strengths, and cross sections for the dielectronic recombination (DR) of $\text{Y}^{+}$, $\text{Sr}^{+}$, $\text{Te}^{2+}$, and $\text{Ce}^{2+}$—low-charge ions relevant to kilonovae and non-local thermodynamic equilibrium (non-LTE) plasmas. Using relativistic atomic structure methods, we computed DR rate coefficients under conditions typical of these environments. These DR rate coefficients and cross sections were calculated using the Flexible Atomic Code (FAC). The DR resonance features were identified by comparing theoretical resonance energies, estimated as the difference between NIST (National Institute of Standards and Technology) excitation energies and Dirac binding energies, with dominant autoionizing states confirmed through analysis of autoionization rates. Our results highlight the critical role of low-lying DR resonances in shaping rate coefficients at kilonova temperatures ($\sim 10^4$ K) and regulating charge-state distributions. Pronounced near-threshold DR resonances significantly influence the evolving ionization states and opacity of neutron star merger ejecta. Comparisons with previous studies emphasize the necessity of including high-$n$ Rydberg states for accurate DR rate coefficients, especially for complex heavy ions with dense energy levels. Discrepancies with existing datasets underscore the need for refined computational techniques to minimize uncertainties. These results provide essential input for interpreting spectroscopic observations of neutron star mergers, including James Webb Space Telescope data. We also put forward suitable candidates for experimental studies, recognizing the challenges involved in such measurements. The data presented here have the potential to refine models of heavy-element nucleosynthesis, enhance plasma simulation accuracy, and improve non-LTE plasma modeling in astrophysical and laboratory settings.

	\end{abstract}
	
	\date{\today}
	
	
	\maketitle

	\section{Introduction}\label{sect:intro}
	
	Dielectronic recombination (DR) affects the ionization balance, radiative cooling, and energy distribution in plasmas, particularly under non-local thermodynamic equilibrium (non-LTE) conditions (\citealp{burgess1964delectronic,gau1980dielectronic,fano1968spectral,rosmej2020dielectronic,bautista2007dielectronic,fritzsche2021dielectronic,badnell2003dielectronic,badnell2006dielectronic,schmidt2008electron}). In such non-LTE environments, where ionization states are not governed by local thermodynamic equilibrium, the interplay between ionization and recombination processes like DR is critical for determining plasma dynamics across both astrophysical and laboratory systems with diverse ionization states. An accurate treatment of DR is crucial for improving plasma models and interpreting observational data.
	
	In laboratory plasmas, DR significantly influences energy loss mechanisms, and enhances plasma stability, particularly in fusion research. The sensitivity of high Rydberg states to weak electric fields enhances DR rates at low plasma densities, making it relevant to environments like the solar corona and magnetic confinement fusion devices (\citealp{rosmej2020dielectronic}). By emitting photons, DR helps regulate plasma temperature, a key factor for both experimental investigations and theoretical models.
	
	Furthermore, DR is essential to collisional-radiative models, which describe the interactions between charged particles and photons in non-LTE plasmas (\citealp{leontyev2016statistical,chung2013comparison}). These models predict plasma responses to varying conditions, including external field effects on ionization potentials and recombination rates. Incorporating DR in these models provides insight into Auger electron heating, hot electron instabilities, and ionization potential depression, which are key to plasma diagnostics and non-equilibrium plasma theory
	(\citealp{galtier2011decay,petitdemange2013dielectronic,rosmej2021plasma}).
	
	DR studies on kilonova-relevant ions remain an area of active investigation. Astrophysical plasmas in kilonovae present extreme, non-LTE environments where ionization structure and radiative processes are strongly influenced. DR plays a dominant role in electron-ion recombination, significantly influencing spectral evolution. Involving high Rydberg states, DR is highly sensitive to density effects, which govern redistributive collisions before radiative stabilization occurs. In kilonovae, which serve as key sites for rapid neutron-capture (r-process) nucleosynthesis at temperatures of approximately $10^4$~K, DR often competes with or even surpasses radiative recombination (RR) in shaping the ionization structure of the ejecta. Under such extreme conditions, DR becomes the predominant recombination mechanism for many heavy ions, surpassing RR (\citealp{mnras2022validity}). This underscores DR’s critical role in determining kilonova ejecta’s ionization structure, transient emission, and the synthesis of elements much heavier than Fe (\citealp{drake2023springer,dalgarno2006applications,eichler1989nucleosynthesis}).
	
	As a kilonova evolves from the diffusion to the nebular phase, recombination alters ionization fractions, changing spectral line intensities and causing the emergence or suppression of features in the optical and near-infrared regions. DR rate coefficients in r-process elements exhibit strong temperature dependence, particularly for singly to doubly ionized species at $T \sim 10^4$~K. Even slight variations in ejecta conditions can significantly affect ionization balance and, consequently, kilonova spectra. In contrast, RR tends to dominate in lower-temperature environments, such as those in the nebular phase of supernovae. Thus, accurate temperature-dependent DR rates are essential for reliable spectral modeling and interpreting transient astrophysical events.  
	
	The importance of DR in kilonovae also stems from the complex electron shell structures of heavy elements, which have dense energy levels and numerous autoionizing states, enhancing DR efficiency. The photon emission in DR during radiative stabilization, serves as a crucial diagnostic tool for interpreting kilonova spectra and constraining plasma parameters such as temperature and electron density. As a result, DR-driven emission plays a vital role in X-ray spectroscopy and plasma diagnostics, offering valuable insights into the physical conditions of astrophysical plasmas (\citealp{rosmej2020dielectronic, song2005dielectronic}).  
	
	Low-charged heavy ions such as \(\text{Y}^{+}, \text{Sr}^{+}, \text{Te}^{2+}, \text{Ce}^{2+}\) (\citealp{Nature2019Sr,Nature2024Te,domoto2022lanthanide}) are highly relevant to kilonovae. 	The James Webb Space Telescope has detected Te$^{2+}$ in GRB 230307A, and Sr$^+$ was detected in kilonova AT2017gfo, underscoring the need for accurate recombination data to understand low-charge ions in astrophysical plasmas (\citealp{Nature2024Te,Nature2019Sr}). Heavy ions with 
	low-lying excited states (e.g., Y$^+$, Sr$^+$, and Ce$^{2+}$) exhibit complex DR behavior even at low densities, causing discrepancies in ionization balance calculations. Moreover, different elements affect kilonova opacity and spectra uniquely: Sr and Y produce strong optical and near-infrared transitions, making them observable, while Te and Ce, with their intricate electronic structures, significantly affect mid-infrared opacities. Their recombination rates determine ionization fractions over time, shaping kilonova spectral evolution. Thus, precise recombination data are essential for improving kilonova models and interpreting spectral signatures accurately.
	
	Accurate data on DR rates, oscillator strengths, and electron-impact cross-sections are essential for quantitative analyses of astrophysical spectra, particularly in kilonova ejecta and non-LTE plasmas. However, the scarcity of DR data for key r-process elements like \(\text{Y}\), \(\text{Sr}\), \(\text{Te}\), and \(\text{Ce}\) introduces uncertainties in modeling ionization balance and spectral evolution. The dense energy level structures of heavy elements further complicate direct comparisons, as recombination rates can vary significantly across ions and even between transitions within the same species. Due to experimental challenges in measuring DR rates for low-charged heavy elements, models often rely on extrapolations from lighter elements (\citealp{MNRAS2023nlte}) or simplified approximations, leading to substantial inaccuracies. In particular, kilonovae and non-LTE plasma models often depend on crude estimates for low-charged heavy ions, where discrepancies can reach an order of magnitude, directly impacting astrophysical predictions. Theoretical studies of these ions are also difficult: while highly charged heavy ions have been widely studied (\citealp{indelicato2005dielectronic,schippers2025testing,schippers2024breit,lindroth2012recombination,lindroth2006determination,Harman2019,beilmann2011prominent,Gonzalez2005,fritzsche2025dielectronic}), research on low-charge species remains limited. To address this gap, the present work uses state-of-the-art atomic structure calculations to provide reliable DR rate coefficients for such ions. These theoretical benchmarks are essential for improving simulation accuracy, refining ionization balance calculations, and enhancing the interpretation of observational data from neutron star mergers and other transient astrophysical events.
	
	To the best of our knowledge, this work presents one of the first computations for the current targets. No experimental measurements exist in the literature, and the only available theoretical data come from a recent study by Banerjee et al. \cite{banerjee2025nebular}, which includes only a few of the ions considered here. Another key objective is to provide reference data for experimentalists to facilitate state-of-the-art research in this field. Recent experimental advancements have enabled the study of low-charge heavy ions at the cryogenic storage ring of the Max Planck Institute for Nuclear Physics in Heidelberg (\citealp{isberner2025electron}). Notably, successful DR measurements on Xe$^{3+}$ represent the highest mass-to-charge ratio experiments conducted to date. However, these experiments face major challenges, including difficulties in ion source generation, high background noise, and weak DR signals that are often obscured (\citealp{isberner2025electron}). Given these experimental limitations, theoretical calculations such as those presented here are required. They provide critical insights and benchmarks not only for experimentalists striving to overcome these challenges but also for astrophysicists modeling and interpreting complex astrophysical phenomena with greater accuracy.

	Section \ref{sect:theory} of this article covers the theoretical specifics that are employed in the present work. The computational results are discussed and presented in Section \ref{sect:results}, while Section \ref{sect:conclusion} contains some concluding remarks.
	
	\section{Theoretical calculations}\label{sect:theory}
	%
	%
	The three DR reactions studied in the present work are:
	
	(i) for Y II to Y I, (for $l \leq 9$ and $n = 6,\dots,35$)
	\begin{align} \label{eq:Y+excitations}
		{\rm Y}^{+}[5s^2 (i)] + e^- &\to \left\{
		\begin{array}{l}
			{\rm Y}^*[5s ~5p ~nl_j (d)] \\ 
			{\rm Y}^*[5s ~4d ~nl_j (d)]
		\end{array} \right\}
		\to {\rm Y}(f) + h\nu\,,
	\end{align}
	
	(ii) for Sr II to Sr I, (for $l \leq 9$ and $n = 6,\dots,35$)
	
	\begin{align} \label{eq:Sr+excitations}
		{\rm Sr}^{+}[5s (i)] + e^- &\to \left\{
		\begin{array}{l}
			{\rm Sr}^*[5p ~nl_j (d)] \\ 
			{\rm Sr}^*[4d ~nl_j (d)]
		\end{array} \right\}
		\to {\rm Sr}(f) + h\nu\,,
	\end{align}
	
	(iii) for Te III to Te II, (for $l \leq 9$ and $n = 6,\dots,35$)
	
	\begin{align}  \label{eq:Te2+excitations}
		{\rm Te}^{2+}[5s^2 ~5p^2 (i)] + e^- &\to \left\{
		\begin{array}{l}
			{\rm Te}^{+*}[5s^2 ~5p ~5d ~nl_j (d)] \\ 
			{\rm Te}^{+*}[5s ~5p^3 ~nl_j (d)]
		\end{array} \right\} \nonumber \\
		&\to {\rm Te}^{+}(f) + h\nu\,,
	\end{align}
	
	(iv) for Ce III to Ce II, (for $l \leq 9$ and $n = 6,\dots,24$)
	
	\begin{align}  \label{eq:Ce2+excitations}
		{\rm Ce}^{2+}[4f^2 (i)] + e^- &\to \left\{
		\begin{array}{l}
			{\rm Ce}^{+^*}[4f~5d~nl_j (d)] 
		\end{array} \right\} \nonumber \\
		&\to {\rm Ce}^{+}(f) + h\nu \,.
	\end{align}

	Here, \( i \) denotes the ground state, \( d \) represent the intermediate states, and \( f \) indexes the final states. The Rydberg electron is represented as \( nl_j \), where \( n \) is the principal quantum number of the captured electron. Here, \( n \) is determined based on convergence. Including more \( n \)-values (e.g., in the hundreds) would have some minor effect on the results, but the difference should not be very significant compared to the values used in the present case. This contributes to some uncertainty in the calculation, which has been accounted for when addressing the overall uncertainty. The emitted decay photons are denoted by \( h\nu \). For a given \( n \), angular momentum states $l = 0, 1, \dots, 9$ and $j = | l \pm \frac{1}{2} | $ are included. Radiative decay involves all electric dipole transitions to lower-lying states resulting in a significant number of states to consider. Except for the \( 4d \) state of Sr\(^+\) and \( 4f \) state of Ce\(^{2+}\), all transitions occur within the same shell (\( \Delta n = 0 \)), while the \( 5s \to 4d \) transition in Sr\(^+\) and \( 4f \to 5d \) transition in Ce\(^{2+}\) involves inter-shell excitation (\( \Delta n = 1 \)).

	For a dielectronic recombination channel, i.e. for a two-step transition $i \to d \to f$, the cross section is expressed as a function of the electron kinetic energy $E$ in the independent resonances approximation as (see, e.g.~\citealp{HaanJacobs,Zim90,Zimmermann,Shabaev1994,Harman2019})
	\begin{equation}
		\label{eq:drkompakt}
		\sigma^{{\mathrm{DR}}}_{i \to d \to f}(E) =
		\frac{2\pi^2}{p^2}  V_a^{i\to d} \frac{A_r^{d \to f}}{\Gamma_d}  L_d(E).
	\end{equation}
	The initial state of the DR process which consists of the ground-state ion and a continuum electron with an asymptotic momentum $\vec{p}$ and spin projection $m_s$. In addition, $\Gamma_d$ is the total natural width of the intermediate autoionizing state, which is the sum of the radiative and autoionization widths: $\Gamma_d = A^d_r + A^d_a$ (here in atomic units with $\hbar=1$). $L_d(E)$ is the Lorentzian line shape function, expressed as
	\begin{equation}
		\label{eq:lorentzian}
		L_d(E) =  \frac{\Gamma_d/(2\pi)}
		{(E_i+E-E_d)^2 +\frac{\Gamma_d^2}{4}},
	\end{equation}
	and is normalized to unity on the energy scale where $p=|\vec{p}|= \sqrt{(E/c)^2 - c^2}$ is the modulus of the free-electron momentum associated with the kinetic energy $E$.
	
	The dielectronic capture rate is related to the rate of its time-reversed process, i.e., the Auger process, by the principle of detailed balance:
	\begin{equation} 
		\label{balance}
		V_a^{i \to d} = \frac{2J_{d}+1}{2(2J_i+1)} A_a^{i \to d} \,.
	\end{equation}
	Here, $J_d$ and $J_i$ are the total angular momenta of the intermediate and the initial states of the recombination process, respectively. Neglecting the energy-dependence of the electron momentum in the vicinity of the resonance, the dielectronic resonance strength, defined as the integrated cross section for a given resonance peak,
	\begin{equation} 
		S^{\mathrm{DR}}_{i \to d \to f} \equiv \int \sigma^{{\mathrm{DR}}}_{i \to d \to f}(E) dE\,,
	\end{equation}
	is given as
	\begin{equation} 
		S^{\mathrm{DR}}_{i \to d \to f}
		= \frac{2\pi^2}{p^2} \frac{1}{2} \frac{2J_{d}+1}{2J_i+1} \frac{A_a^{i \to d}A_r^{d \to f}}{A_r^{d}+A_a^{d}}\,,
		\label{eq:strength}
	\end{equation}
	where $A_a^{i \to d}$ is implicitly defined in Eq.~\eqref{balance}. The factor $\frac{2\pi^2}{p^2}$ defines the phase space density and the $1/2$ stems from the spin degeneracy of the free electron.
	
	The total rate coefficients ($\alpha_{DR}$) for astrophysical and thermal plasmas are described by
	\begin{align} \label{rate-coeff-eq}
		\alpha_{DR}(T) &= \frac{h^3}{(2 \pi m_e k T)^{3/2}} 
		\sum_{d} \frac{2J_{d}+1}{2 (2J_{i}+1)} \nonumber \\
		&\quad \times \frac{A_a^{i \to d} A_r^{d \to f}}{A_r^{d} + A_a^{d}} 
		\exp \left(-\frac{E}{k T} \right)\,,
	\end{align}
	derived by summing across all possible autoionization channels and averaging over the Maxwellian distribution of electron energies (\citealp{dubau1980dielectronic}). In this expression, $k$ denotes the Boltzmann constant, $h$ is the Planck constant that we write explicitly here, $T$ represents the electron temperature, and $E$ is the resonance energy.

	\section{Results and discussion}\label{sect:results}
	
	In this study, the relativistic configuration interaction method with independent-particle basis wave functions was utilized to calculate the energy levels, radiative rates, autoionization rates, and DR cross sections and rates as implemented in the Flexible Atomic Code (FAC) (\citealp{FAC2003ApJGu,FACgu2004AIP,singh2025dielectronic}). The relativistic distorted-wave approximation was employed to describe the continuum states. For the calculation of wave functions and energy levels corresponding to the initial states, intermediate doubly excited states, and radiative final states, contributions from electron correlations, quantum electrodynamics (QED) effects, and Breit interactions were systematically accounted for, ensuring accurate determination of energy levels and wave functions.

	The DR resonance features were identified by comparing the theoretically calculated resonance energies with known excitation energies of the core ion, as listed in the NIST Atomic Spectra Database. The DR resonance energy, $E_{\text{DR}}$, is approximately given by the difference between the excitation energy of the bound core electron, $E_{\text{ex}}$, and the binding energy of the captured free electron, $E_{\text{B}}$; that is, E$_{DR}$ = E$_{ex}$ - E$_B$. Here, $E_{\text{ex}}$ values were taken from NIST, while $E_{\text{B}}$ was estimated using Dirac binding energies (\citealp{bernhardt2015electron}). This provides an approximate position for the expected DR resonance features. To further verify these identifications, we analyzed the autoionization rates of the intermediate states. Since higher autoionization rates typically correspond to stronger DR resonance features, the dominant contributors identified through rate analysis were found to be consistent with the states inferred from energy comparisons.

	\begin{figure}
		\begin{center}		
			\includegraphics[width=0.6\textwidth, trim=2cm 6cm 3cm 0cm]{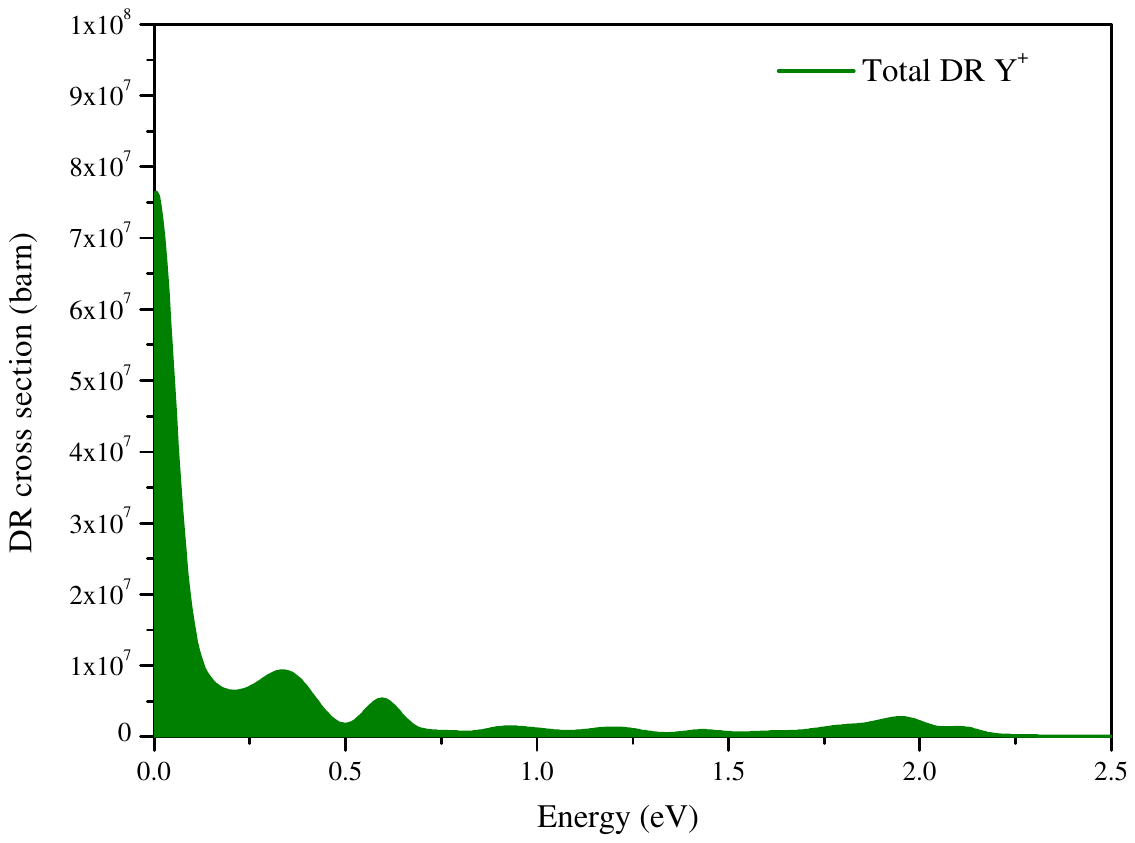}
			\caption{Total DR cross section for Y$^{+}$ recombining into Y is plotted against the relative electron energy. The Lorentzian line shapes are convoluted with a Gaussian function with a width of 100-meV.}
			\label{Y+DR}
		\end{center}
	\end{figure}
	
	Figure~\ref{Y+DR} illustrates the total DR cross sections for the recombination of Y$^{+}$ into Y. The resolution width is set to 100~meV, modeled as a Lorentzian profile convoluted with a Gaussian. The calculations include Rydberg states up to $n = 6, \ldots, 35$. A strong DR resonance is observed near the threshold, accompanied by relatively weaker resonance peaks at approximately 0.3~eV and 0.6~eV. These peaks are attributed to the $4d5s~^3D_J$ states (where $J = 1, 2, 3$). Additionally, a small resonance around 2~eV is associated with the $5s5p~^3P^o_J$ states (where $J = 0, 1, 2$). 
	
	The DR resonance strength for Y$^{+}$ is approximately an order of magnitude higher than that of previous ions studied by our group (\citealp{singh2025dielectronic,isberner2025electron}). The presence of strong resonances at low energies suggests that DR is dominated by optically allowed core excitations, and it is a dominant recombination mechanism for Y$^{+}$ under various plasma conditions, particularly in astrophysical environments. Given the exceptionally strong resonance near the threshold, Y$^{+}$ is an excellent candidate for experimental validation, as background effects are expected to be minimal, enhancing detection feasibility. However, experimental challenges remain, particularly in distinguishing the DR signal from radiative recombination, which is typically prominent at near-threshold energies. Effectively separating these contributions will be crucial for future measurements. The findings of this work could facilitate both observational and experimental studies of Y$^{+}$ in the near future.

	\begin{figure}
		\begin{center}		
			\includegraphics[width=0.6\textwidth, trim=2cm 6cm 3cm 0cm]{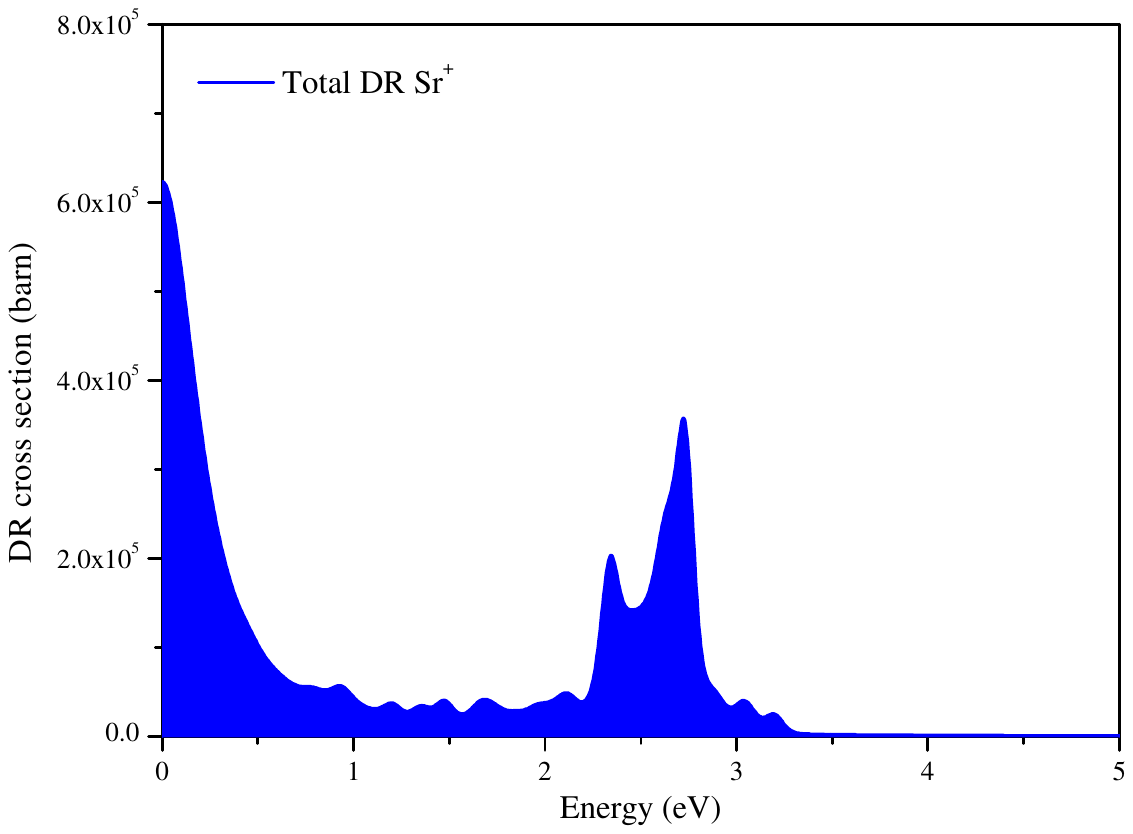}
			\caption{Total DR cross section for Sr$^{+}$ recombining into Sr is plotted against the relative electron energy.}
			\label{Sr+DR}
		\end{center}
	\end{figure}

	The total DR spectrum for Sr$^{+}$ recombining to neutral Sr is presented in Figure~\ref{Sr+DR}, with a resolution width set to 100~meV. The DR cross section exhibits characteristic resonance peaks, arising from the capture of free electrons into autoionizing Rydberg states via core excitations ($5s \to 5p, 4d$). A strong DR resonance feature is observed near zero energy; however, its exact origin remains unidentified. There is a strong possibility that it results from a theoretical artifact associated with the Rydberg state $n = 6$, as observed in our calculations, since no such excited states have been reported in the literature at such low energies. Another prominent resonance appears at approximately 2.3~eV, primarily originating from the $4p^6 4d~^2D_{5/2}$ and $^2D_{3/2}$ states, with the $^2D_{5/2}$ state being the dominant contributor. Additionally, a strong DR resonance is observed around 2.7~eV, attributed to the $4p^6 5p~^2P^o_{1/2}$ and $^2P^o_{3/2}$ states. These states are also responsible for the knee-like resonances observed around 3~eV.  
	
	Among the ions studied in this work, the DR spectrum of Sr$^{+}$ exhibits the weakest signal strength, possibly presenting significant challenges for experimental observation with currently available setups. Its DR signal strength is approximately one to two orders of magnitude smaller than that of the other ions studied in this work as well as in previous works (\citealp{singh2025dielectronic,isberner2025electron}). To detect this spectrum experimentally, it would be necessary to significantly reduce background noise.

	\begin{figure}
		\begin{center}		
			\includegraphics[width=0.6\textwidth, trim=2cm 6cm 3cm 0cm]{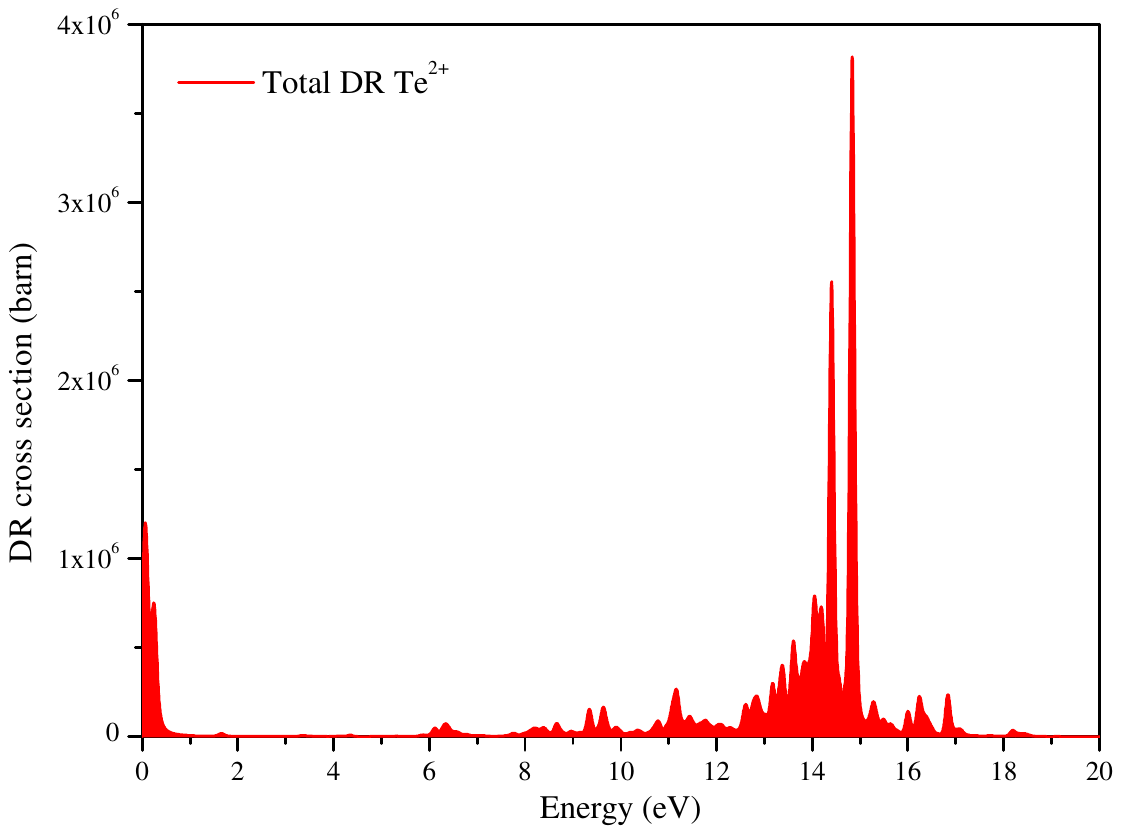}
			\caption{Total DR cross section for Te$^{2+}$ recombining into Te$^{+}$ is plotted against the relative electron energy.}
			\label{Te2+DR}
		\end{center}
	\end{figure}
	
	Figure~\ref{Te2+DR} presents the total DR cross section for Te$^{2+}$ recombining into Te$^{+}$, calculated with a resolution width of 100~meV. In the DR spectrum, a small resonance is observed at low energies, below 0.5~eV, attributed to the $5s^2 5p^2~^3P_J$ states (where $J = 0, 1, 2$). At approximately 14.5~eV, two prominent resonance structures appear, corresponding to the $5p 5d~^3D^o_J$ (where $J = 1, 2, 3$) and $^3P^o_J$ (where $J = 0, 1, 2$) states, with the $^3D^o_1$ and $^3P^o_1$ states being the dominant contributors. Additionally, several significant resonance peaks are observed on either side of this strong resonance feature, arising from the $5s 5p^3$ states.  
	
	Given the presence of a well-defined resonance near the threshold and strong DR resonance features at higher energies, along with a relatively high DR signal strength, this ion is a promising candidate for further DR spectroscopy studies, both experimentally and theoretically.

	\begin{figure}
		\begin{center}		
			\includegraphics[width=0.6\textwidth, trim=2cm 6cm 3cm 0cm]{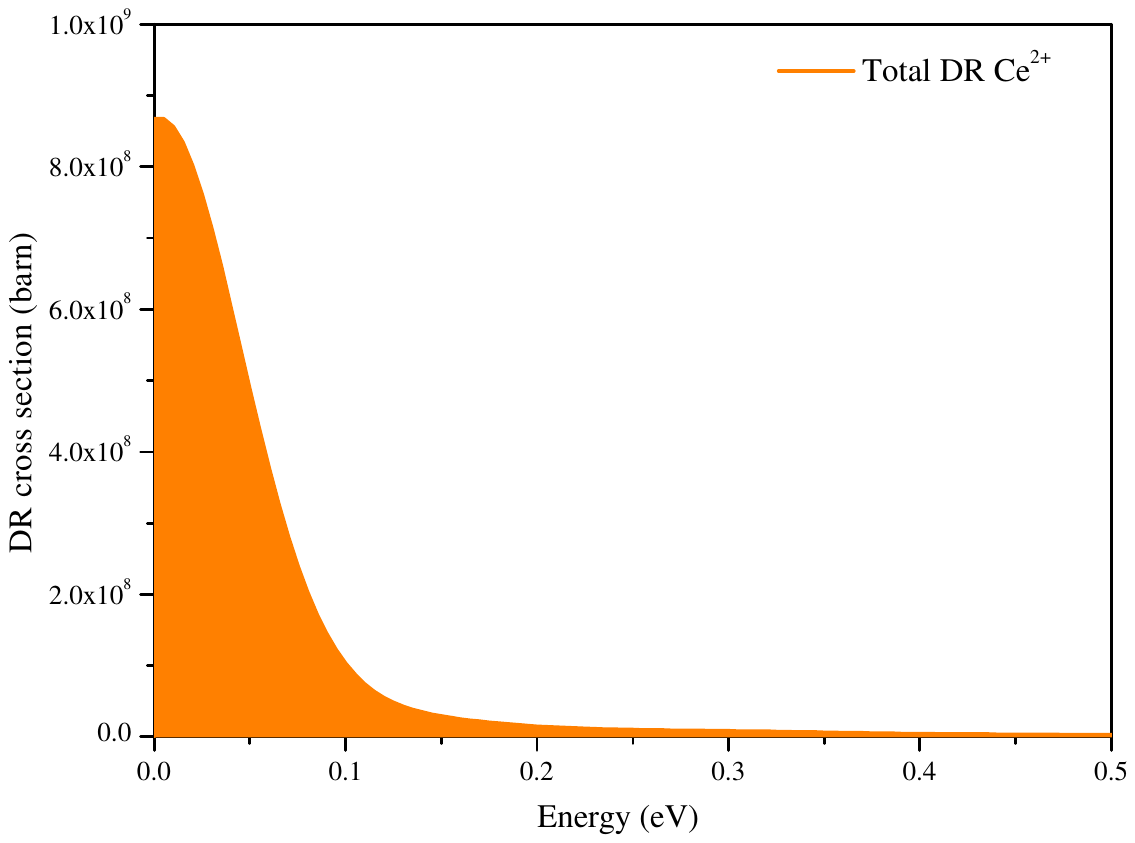}
			\caption{Total DR cross section for Ce$^{2+}$ recombining into Ce$^{+}$ is plotted against the relative electron energy.}
			\label{Ce2+DR}
		\end{center}
	\end{figure}

	Figure~\ref{Ce2+DR} presents the total dielectronic recombination (DR) cross sections for the recombination of Ce$^{2+}$ into Ce$^{+}$, calculated with a resolution width of 100~meV. A strong DR resonance is observed near the threshold, primarily due to the convolution of the $4f^2$ and $4f~5d$ states. At energies close to 0~eV, the $4f^2~^3H_5$ state is the dominant contributor, whereas toward the tail end of the peak, the $4f5d^1G^o_4$ state becomes the most significant. As seen in Table~\ref{tab:fac-vs-nist}, although the excitation energy of the $4f5d^1G^o_4$ state appears lower than that of the $4f^2~^3H_5$ state in the FAC calculations, the inclusion of the Rydberg electron introduces additional configuration mixing. As a result, the resonance energy associated with the $4f5d^1G^o_4$ state becomes higher than that of the $4f^2~^3H_5$ state. The contribution of the $4f5d^1G^o_4$ state at the tail end of the resonance feature is further supported by the analysis of the autoionization rates. Although the spectrum appears to decline beyond 0.2~eV, the cross-section magnitude remains on the order of $10^7$ barns, emphasizing the strength of this prominent resonance feature. This suggests that Ce$^{2+}$ could serve as a strong candidate for experimental study near the threshold energy, similar to Y$^{+}$. Furthermore, the exceptionally high DR signal might indicate a considerable probability of its detection in a kilonova.

	\begin{table}[ht]
		\centering
		\caption{Comparison of energy levels (in eV) from present FAC calculations and NIST (\citealp{NIST_ASD}) for selected ions.}
		\label{tab:fac-vs-nist}
		\begin{tabular}{|l|l|c|c|}
			\hline
			Ion & State & FAC  & NIST  \\
			&  &  & (\citealp{NIST_ASD}) \\
			\hline
			\multirow{9}{*}{Y$^+$} 
			& 4d5s$^2$ $^2$D$_{3/2}$ (Y) & 0 & 0 \\
			& 5s$^2$ $^1$S$_0$       & 4.728 & 6.217 \\
			& 4d5s $^3$D$_1$         & 4.978 & 6.321 \\
			& 4d5s $^3$D$_2$         & 4.994 & 6.347 \\
			& 4d5s $^3$D$_3$         & 5.021 & 6.397 \\
			& 5s5p $^3$P$^\circ_0$   & 7.626 & 9.124 \\
			& 5s5p $^3$P$^\circ_1$   & 7.637 & 9.165 \\
			& 5s5p $^3$P$^\circ_2$   & 7.638 & 9.273 \\
			\hline
			\multirow{6}{*}{Sr$^+$} 
			& 5s$^2$ $^1$S$_0$ (Sr)      & 0 & 0 \\
			& 4p$^6$5s $^2$S$_{1/2}$ & 4.603 & 5.695 \\
			& 4p$^6$4d $^2$D$_{3/2}$ & 6.878 & 7.500 \\
			& 4p$^6$4d $^2$D$_{5/2}$ & 6.879 & 7.534 \\
			& 4p$^6$5p $^2$P$^\circ_{1/2}$ & 7.279 & 8.635 \\
			& 4p$^6$5p $^2$P$^\circ_{3/2}$ & 7.360 & 8.735 \\
			\hline
			\multirow{10}{*}{Te$^{2+}$}
			& 5s$^2$5p$^3$ $^4$S$^\circ_{3/2}$ (Te$^{+}$) & 0 & 0 \\
			& 5s$^2$5p$^2$ $^3$P$_0$          & 17.798 & 18.600 \\
			& 5s$^2$5p$^2$ $^3$P$_1$          & 18.274 & 19.189 \\
			& 5s$^2$5p$^2$ $^3$P$_2$          & 18.740 & 19.612 \\
			& 5s$^2$5p5d $^3$D$^\circ_1$      & 31.560 & 32.950 \\
			& 5s$^2$5p5d $^3$D$^\circ_3$      & 32.466 & 33.589 \\
			& 5s$^2$5p5d $^3$D$^\circ_2$      & 31.872 & 33.789 \\
			& 5s$^2$5p5d $^3$P$^\circ_0$      & 32.581 & -- \\
			& 5s$^2$5p5d $^3$P$^\circ_2$      & 32.665 & 33.070 \\
			& 5s$^2$5p5d $^3$P$^\circ_1$      & 32.628 & 33.204 \\
			\hline
			\multirow{5}{*}{Ce$^{2+}$}
			& 4f5d$^2$ $^4$H$^\circ_{7/2}$ (Ce$^{+}$)                         & 0 & 0 \\
			& 4f$^2$ $^3$H$_4$                             & 12.388 & 10.956 \\
			& 4f$^2$ $^3$H$_5$                             & 12.498 & 11.145 \\
			& 4f$^2$ $^3$H$_6$                             & 12.642 & 11.344 \\
			& 4f5d $^1$G$^\circ_4$                         & 10.701 & 11.362 \\
			\hline
		\end{tabular}
	\end{table}

	Table~\ref{tab:fac-vs-nist} presents a comparison between the energy levels calculated in this work using the FAC and the values available from the NIST Atomic Spectra Database~(\citealp{NIST_ASD}). As noted by \cite{FAC2003ApJGu}, energy levels computed using FAC can exhibit uncertainties due to incomplete treatment of correlations or limitations in the atomic model or lack of configuration mixing. From Table~\ref{tab:fac-vs-nist}, it is evident that the calculated energy levels deviate from the NIST values by approximately 1 to 1.5 eV. Such differences are expected for near-neutral heavy ions, as the estimated uncertainty in energy levels calculated using FAC can be of the order of a few electronvolts. A recent study by \cite{nahar2024theoretical}, which also investigates low-charged lanthanides relevant to kilonovae, reports similar discrepancies of a few eVs when comparing theoretical values with NIST data. This supports the observation that such deviations are typical when modeling complex atomic structures. In addition to energy uncertainties, there are inherent uncertainties in the computed rates and cross sections. According to the FAC manual authored by \cite{Gu_FAC}, in the case of near-neutral ions (as in the present case), the uncertainties associated with radiative decay rates and autoionization rates can reach 20\% or, in certain instances, exceed this value. Additionally, limiting the number of states \( n \) introduces an uncertainty of approximately 10\%. Therefore, the overall uncertainty in the present calculation of DR cross section and strengths is estimated to be around 30\%. Given the substantial amount of data generated in this work, all data are made available as supplementary material.

	\begin{figure}
		\begin{center}		
			\includegraphics[width=0.6\textwidth, trim=2cm 6cm 3cm 0cm]{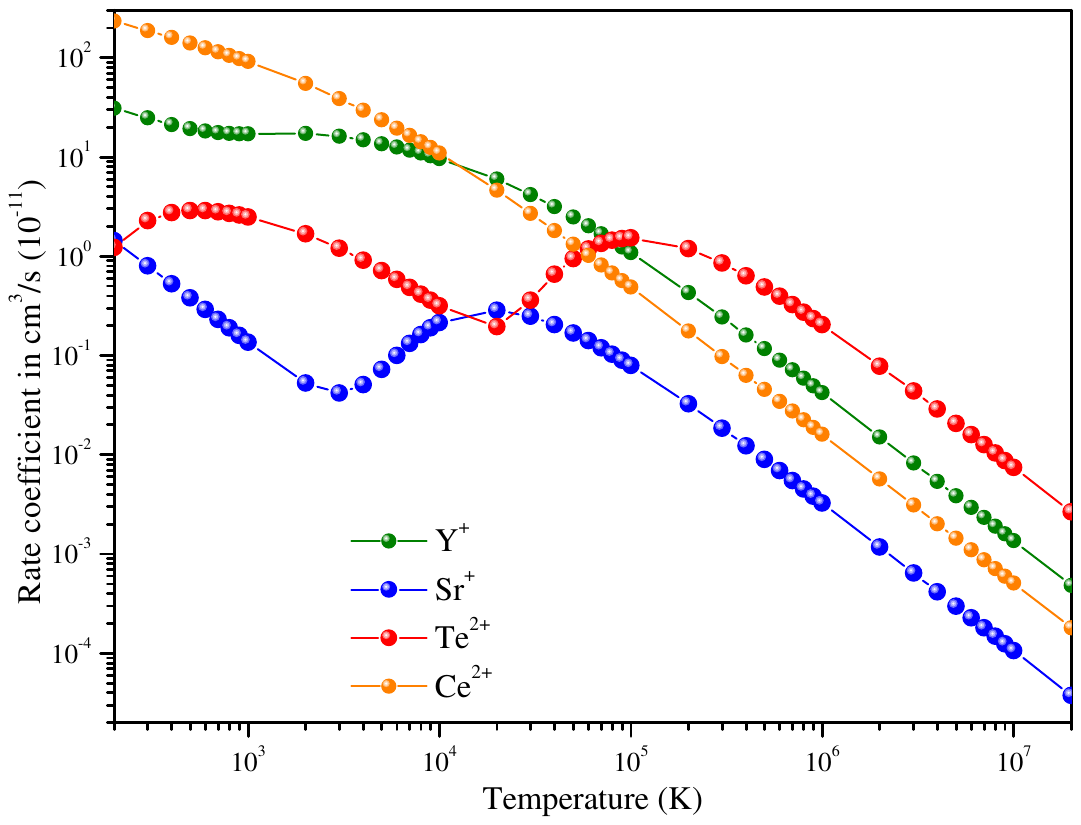}
			\caption{Rate-coefficient for Y$^{+}$, Sr$^{+}$, Te$^{2+}$, and Ce$^{2+}$ is plotted against the electron temperature.}
			\label{Rate}
		\end{center}
	\end{figure}

	Figure~\ref{Rate} illustrates the dependence of the total DR rate coefficients on electron temperature, calculated by incorporating Rydberg states with principal quantum numbers ranging from $n = 6$ to $n = 35$, except for $\text{Ce}^{2+}$. While the high-temperature rates ($T > 10^5$ K) may not have significant practical relevance, they are included for completeness. At high temperatures $T$, the DR rate coefficient decreases following a $\sim T^{-3/2}$ dependence, as the increasing kinetic energy of free electrons reduces the probability of recombination. In contrast, at low temperatures, the DR process is predominantly influenced by a number of resonances situated just above the threshold. These resonances contribute significantly to the recombination rate, ensuring that DR remains efficient even at lower temperatures. This occurs because the $\sim T^{-3/2}$ factor compensates for the lowness of the temperature. The calculated values have an estimated uncertainty of around 38\%, determined using the error propagation rule from Eq. \ref{rate-coeff-eq}. At low temperatures, DR is primarily governed by near-zero-energy resonances, where the energy of free electrons closely matches the energy levels of states within the ion, enabling highly efficient recombination processes. The dominance of these resonances enhances the recombination probability, significantly influencing the ionization balance of the kilonova ejecta. Notably, the behavior of DR at these temperatures can resemble radiative recombination, highlighting the need for precise resonance data for accurate modeling. As the temperature increases, contributions from high-$n$ Rydberg states become more prominent. These states, with their highly excited electrons near the ionization threshold, play a crucial role in recombination dynamics at higher temperatures. The interplay between the thermal energy of the electrons and the populations of these excited states alters the overall recombination efficiency, resulting in a shift in the effective ionization balance. This shift is particularly significant for ions with one or two electrons outside closed shells (e.g., $\text{Y}^{+}$, $\text{Sr}^{+}$, and $\text{Ce}^{2+}$), as it impacts the interpretation of kilonova spectra and the modeling of non-LTE plasmas (\citealp{badnell2003dielectronic}).
	
	It is worth noting that the uncertainty reported in the present work may appear large; however, we have included an additional 10\% uncertainty to account for the contribution from higher Rydberg states that were not included in the present calculations. If experimental observations are carried out for the current set of ions, it is highly likely that these contributions would also be suppressed due to field ionization effects, which typically limit the population of high-$n$ Rydberg states in experimental setups. Therefore, the additional 10\% uncertainty we have included would have a relatively minor impact when comparing with experimental results. Nonetheless, we have included this 10\% additional uncertainty to provide a more cautious upper bound on the theoretical uncertainty. Furthermore, several previous studies on both high- and low-charge ions (\citealp{kaur2018dielectronic,bleda2022dielectronic,nahar2024theoretical}) have reported similarly large uncertainties, both in terms of energy scale and rate coefficient, which is to be expected when dealing with complex, many-electron systems such as those considered in this work.
	
	\begin{table}[h]
		\centering
		\caption{DR rate coefficients (\(10^{-11} \, \text{cm}^3/\text{s}\)) at  temperatures \(10^3\), \(10^4\), and \(10^5\) K.}
		\label{tab:dr_rates}
		\begin{tabular}{lccc}
			\toprule
			Ion & \(10^3\) K & \(10^4\) K & \(10^5\) K \\
			\midrule
			Y\(^+\)     & 17.02138 & 9.61403  & 1.08042 \\
			Sr\(^+\)    &  0.13610 & 0.21363  & 0.07922 \\
			Te\(^{2+}\) &  2.48856 & 0.31646  & 1.52769 \\
			Ce\(^{2+}\) & 91.13522 & 10.96778 & 0.48809 \\
			\bottomrule
		\end{tabular}
	\end{table}
	
	Table~\ref{tab:dr_rates} presents the DR rate coefficients at temperatures of \(10^3\), \(10^4\), and \(10^5\) K. The complete set of DR rate coefficients as a function of temperature is provided in the supplementary material. Among the studied ions, \(\text{Ce}^{2+}\) exhibits relatively high DR rate coefficients at very low temperatures compared to the other ions. In the kilonova-relevant temperature range (\(\sim 10^4\) K), as shown in Table~\ref{tab:dr_rates}, the rate coefficients for \(\text{Y}^+\) and \(\text{Ce}^{2+}\) become comparable and significantly exceed those of \(\text{Sr}^{+}\) and \(\text{Te}^{2+}\). This indicates that \(\text{Y}^+\) and \(\text{Ce}^{2+}\) have a stronger tendency to populate lower ionization states over time, while \(\text{Sr}^{+}\) and \(\text{Te}^{2+}\) remain ionized for a longer duration in the expanding ejecta. These DR trends have critical implications for kilonova spectra. Higher DR rates, as seen for \(\text{Y}^+\) and \(\text{Ce}^{2+}\), are expected to promote efficient recombination to lower ionization states, contributing to optical and near-infrared spectral features. In contrast, the slower recombination associated with \(\text{Sr}^{+}\) and \(\text{Te}^{2+}\) can lead to the retention of higher ionization states, producing spectral features characterized by higher-energy transitions and shifts toward shorter wavelengths. The availability of these computed DR rates will be instrumental in refining spectral models, improving predictions of line emission, and enhancing our understanding of kilonova ejecta composition, as they provide crucial inputs for astrophysical modeling tools like the SUMO spectral synthesis code used in kilonovae simulations (\citealp{jerkstrand201144ti}).

	\begin{figure}
		\begin{center}		
			\includegraphics[width=0.6\textwidth, trim=2cm 6cm 3cm 0cm]{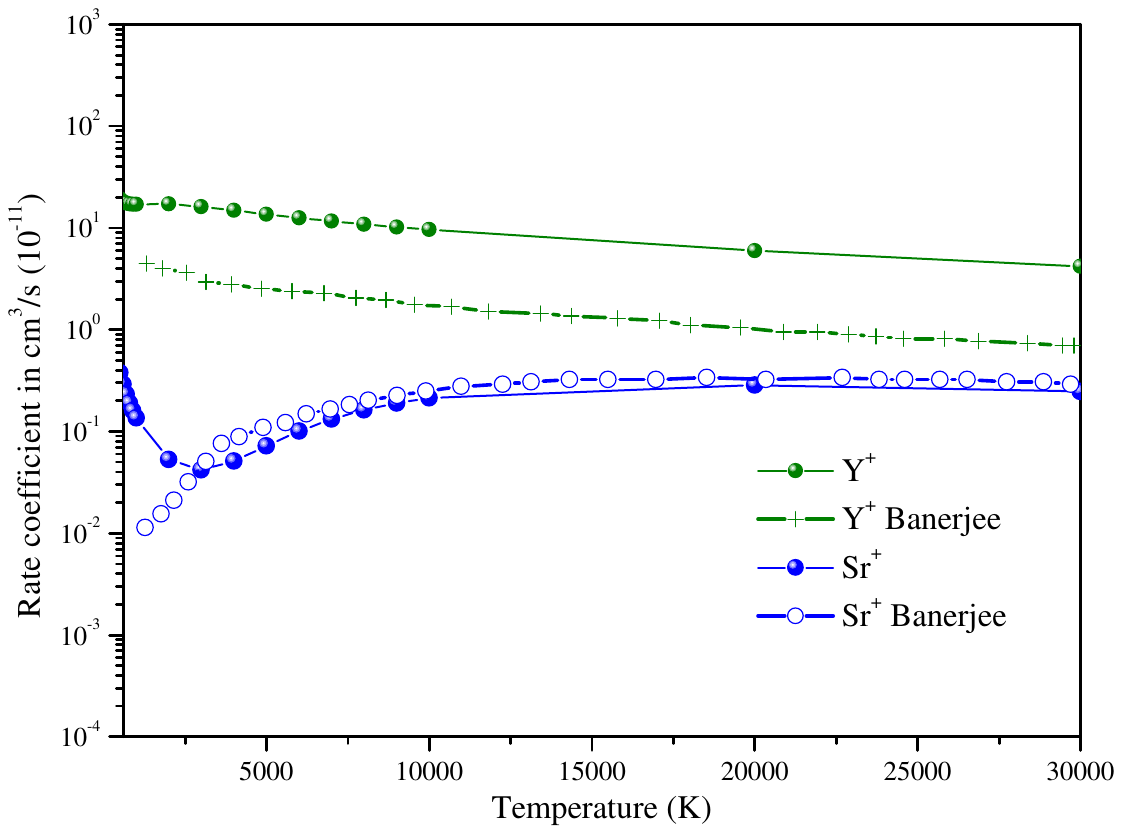}
			\caption{Comparative study for rate-coefficient for Y$^{+}$, and Sr$^{+}$ with available study from Banerjee et al. \cite{banerjee2025nebular}.}
			\label{Rate-comparison}
		\end{center}
	\end{figure}

	In Figure \ref{Rate-comparison}, a comparison is made between the DR rate coefficients obtained in this work and those reported in the only existing study over the temperature range characteristic of the nebular phase of kilonovae. During this phase, the temperature is typically around 10,000 K (\citealp{Metzger2019,tanaka2020systematic}). To account for the temperature evolution from the early weeks after the event to later stages, influenced by the evolving ejecta composition, the rate coefficients are presented across a broader temperature range from 1,000 K to 30,000 K.
	
	In the case of \(\text{Y}^+\), the present values are approximately an order of magnitude larger than those reported by Banerjee et al. \cite{banerjee2025nebular}. The same number of intermediate states has been considered in both studies; however, the difference in magnitude arises from the inclusion of a significantly larger number of Rydberg states in the present work. The inclusion of high-$n$ Rydberg states enhances the recombination probability (\citealp{Nahar1997}). Additionally, differences in the number of orbital angular momentum states (\( l \)) considered in the two studies may further contribute to the observed discrepancy in the rate coefficients. 
	
	A different trend is observed for \(\text{Sr}^+\), where the present results show good agreement with those of Banerjee et al. \cite{banerjee2025nebular}. This consistency can be attributed to the inclusion of an additional \( 4p^6 4f \) state in their study, which increases their rate coefficient. However, this enhancement is counterbalanced by their lower number of Rydberg states and \( l \)-values, resulting in an overall accidental numerical agreement between the two datasets. The role of Rydberg states is particularly significant in singly charged ions like \(\text{Y}^+\) and \(\text{Sr}^+\), where dense energy levels allow for extensive recombination pathways (\citealp{dunn1992early}). Despite some variations in magnitude, the present DR rate coefficients remain within an order of magnitude—or better—compared to previous studies. This level of consistency supports the reliability of the present work.  
	
	\section{Conclusions}\label{sect:conclusion}
	
	Given the importance of DR in kilonova spectra and non-LTE plasmas, systematic calculations of DR rates, strengths, and cross sections for low-charge heavy ions such as $\text{Y}^{+}$, $\text{Sr}^{+}$, $\text{Te}^{2+}$, and $\text{Ce}^{2+}$ are critical. These data are essential for interpreting transient emissions across multiple electromagnetic bands in kilonovae and improving plasma models for astrophysical and laboratory contexts. Studying DR processes in non-LTE plasmas enhances our understanding of plasma behavior across diverse environments, advancing theoretical and experimental plasma physics.
	
	Strong near-threshold DR resonances for $\text{Y}^{+}$ and $\text{Ce}^{2+}$ highlight their role in modifying the ionization balance of kilonova ejecta. DR rate sensitivity to plasma conditions, such as temperature and density, underscores the need for precise data to refine ionization state modeling. These rate coefficients are crucial for interpreting observational data from instruments like the James Webb Space Telescope, as seen in the detection of tellurium in specific kilonovae. Additionally, the methodology employed in this study can be extended to further DR investigations, improving simulations beyond crude approximations currently in practice.
	
	Comparing our calculated DR rates with previous studies reveals the impact of including high-$n$ Rydberg states, illustrating challenges in achieving accurate results for complex heavy ions. Discrepancies between our data and other studies highlight the need for continued advancements in computational techniques to reduce uncertainties in DR rate coefficients. This work identifies several candidates for experimental measurements, such as $\text{Y}^{+}$ and $\text{Ce}^{2+}$ for near-threshold DR studies due to their strong DR signal strengths. $\text{Te}^{2+}$ also emerges as a promising candidate for DR spectrum studies and benchmarking DR analyses. These findings enhance interpretations of transient phenomena and provide a foundation for experimental validation.
	
	\section{Data availability}
	The supplementary data set is available at DOI: 10.5281/zenodo.15683884.

	\bibliography{references}

\end{document}